\pdfoutput=1
\documentclass[twocolumn, aps, prd, groupadress, nofootinbib, showkeys]{revtex4-2}
\usepackage{amsmath, amsfonts, amssymb, mathrsfs, amsthm}

\usepackage{mathpazo}

\usepackage[utf8]{inputenc}

\renewcommand{\date}{}

\usepackage[hyperindex]{hyperref}
\hypersetup{breaklinks=true, colorlinks=true,
allcolors = blue
}

\newtheorem{theorem}{Theorem}[section]

\newtheorem{corollary}[theorem]{Corollary}
\newtheorem{lemma}[theorem]{Lemma}

\theoremstyle{remark}

\theoremstyle{definition}

\usepackage[all]{hypcap}

\begin{document}

\title{\Large{Mass generation and gravity}}

\author{Mario Novello}\email[Electronic address:]{novello@cbpf.br}
 \affiliation{Centro de Estudos Avan\c{c}ados de Cosmologia,
 Rio de Janeiro, RJ, Brazil}
 \affiliation{Centro Brasileiro de Pesquisas Físicas, %Rua Dr. Xavier Sigaud 150, Urca, CEP 22290-180,
 Rio de Janeiro, RJ, Brazil}
\author{Vicente Antunes}\email[Electronic address:]{antunes@cbpf.br}
 \affiliation{Centro de Estudos Avan\c{c}ados de Cosmologia,
 Rio de Janeiro, RJ, Brazil}
 \affiliation{Centro Brasileiro de Pesquisas Físicas, %Rua Dr. Xavier Sigaud 150, Urca, CEP 22290-180,
 Rio de Janeiro, RJ, Brazil}

\begin{abstract}

\begin{center}
Accepted for publication in \textit{Gravitation and Cosmology} 28 (2022)
\end{center}

Despite the success of the Higgs mechanism to account for the generation of the masses of Standard Model (SM) elementary particles, the ultimate nature and origin of ``mass'' remain open questions in contemporary physics. From a foundational perspective, mass should be fundamentally related to the gravitational interaction and, according to Mach, to the structure of the Universe. In the present letter, a fully dynamical mass generation mechanism induced by higher-order corrections in the GR Lagrangian is discussed. Notably, the vacuum energy plays a key role in this process, which applies to both vector bosons and fermions. We show that, at the classical level, the Higgs mechanism can be thought of as a particular case of the present mechanism. 

\end{abstract}

%\pacs{98.80.-k, 98.80.Cq, 04.50.Kd}

\maketitle

\section{Introduction}

The Brout-Englert-Higgs (BEH) mechanism \cite{ANDERSON_1963PhRv_130_439A, ENGLERT_BROUT_1964_PhRvL_13_321E, HIGGS_1964_PhRvL_13_508H, GURALNIK_HAGEN_KIBBLE_1964PhRvL_13_585G} is considered the key ingredient that brought consistency to the Standard Model (SM) of Fundamental Particles by rendering the underlying quantum field theories (QFTs) gauge invariant and renormalizable. The detection of a spinless bosonic particle with a mass of approximately 125 GeV and properties compatible with the SM Higgs boson \cite{ATLAS_CMS_2015_PhRvL_114s_1803A, MURRAY_SHARMA_2015_ARNPS_65_515M} provided strong support for this model. Even though the BEH mechanism is generally regarded as the ``fundamental'' process responsible for the origin of the masses of the SM fundamental particles, the physical mass of the Higgs boson is itself defined by a free parameter of the SM. Moreover, unresolved issues such as fine-tuning and the hierarchy problems of the BEH theory, unanswered questions regarding a possible composite nature of the Higgs boson(s), issues concerning the origin of neutrino masses and the neutrino mass scale, and the cosmological constant problem, all indicate that the SM is far from being a complete picture of Nature.

It is universally recognized that the concept of ``mass'' is poorly understood in contemporary physics.
 First of all, mass is a strictly classical property that was imported to quantum mechanics with no further warrant. 
Second, if mass is indeed a quantized quantity, then the observed mass spectrum of SM requires the corresponding mass operator to describe bound states of more fundamental constituents. By consistency, one would therefore be forced to assume that the ultimate fundamental constituents are massless and that mass must have a strictly dynamical origin. Moreover, one can only hope that the theory capable of implementing this program will be essentially non-perturbative, UV-complete and dispense with the undesirable features of standard QFTs. Although some progress in these particular directions was made by string theories, all attempts to formulate a quantum theory in which the masses of \textit{all} fundamental particles are dynamically generated have so far failed.
 
 On the other hand, the idea that mass should have a strictly dynamic origin (already at the classical level) was first suggested a long time ago by Mach \cite{MACH}. According to some interpretations of the notorious ``Mach's principle'', the mass-inertia of localised systems should somehow be determined by the global structure of the Universe \cite{BONDI_SAMUEL_1997}. A formal, mathematically precise expression of this principle, however, remains elusive. On the other hand, in the light of General Relativity (GR), it is generally understood that mass should be fundamentally related to the gravitational interaction, which is also the dominant influence in the Universe at large scales. It is also generally believed that a more complete theory that supersedes the SM will necessarily incorporate (or be incorporated by) gravity. If we accept that GR is also the low energy limit of some UV-complete theory, then the non-minimal curvature-matter couplings among the higher-order corrections in the corresponding effective Lagrangian suggest a bridge between the quantum realm and the Cosmos at the most fundamental level.
 In that spirit, a mass generation process mediated by gravity through non-minimal curvature-matter couplings was proposed in \cite{NOVELLO_2011_CQGra_28c_5003N, NOVELLO_BITTENCOURT_2012_PhRvD_86f_3510N}, both for vector bosons and for fermions. Here we present a simpler way to achieve that same goal.

\section{A redundant Lagrangian for Maxwell\rq s dynamics}

 Consider a vector field $ W^{\mu}$ coupled to gravity with dynamics defined by the Lagrangian
 \begin{widetext}
 \begin{equation}
 L_{0} = - \frac{1}{4} W_{\mu\nu}  W^{\mu\nu} + \frac{1}{\kappa}  R + V(\Phi) R + \frac{1}{2}  H(\Phi) \Phi^{\mu} \Phi_{\mu} + L_{m} 
 \label{1}
 \end{equation}
 \end{widetext}
where the tensor $ W_{\mu\nu}$ is given in terms of the potential by the formula $ W_ {\mu\nu} = W_{\mu \, ,\nu} - W_{\nu \, ,\mu} $, the scalar $ \Phi $ is the squared norm $ \Phi =  W_{\mu} W^{\mu}$,
 $ L_{m}$ is the matter Lagrangian of the main component responsible for the geometry, $\kappa = 8\pi G$ is the gravitational constant, and the comma means ordinary derivative $\Phi_{\mu} = \partial_{\mu} \Phi = \Phi_{, \mu}$. Here, we will consider only the Abelian case, for simplicity. The dynamical equation for the vector field is
\begin{equation}
W^{\mu\nu}{ }_{;\nu} +  %\frac{2}{c^2}\, 
2\Omega W^{\mu} = 0,
\label{13625}
\end{equation}
where the semicolon means covariant derivative, $W_{\mu;\nu} = \nabla_{\nu}W_{\mu}$, and
 $\Omega(\Phi) = R V' - \frac{1}{2} H' \Phi_{\alpha}\Phi^{\alpha} - H \Box \Phi$, a prime meaning derivative with respect to $ \Phi.$
Variation of $ g_{\mu\nu}$ yields
\begin{widetext}
\begin{equation}
\left(\frac{1}{\kappa} + V\right) \left( R_{\mu\nu} - \frac{1}{2} R g_{\mu\nu} \right) = - E_{\mu\nu} - \tau_{\mu\nu} - T_{\mu\nu}^{m} - \frac{\lambda}{\kappa}g_{\mu\nu}
\label{13618}
\end{equation}
\end{widetext}
where
\begin{equation*}
E_{\mu\nu} = W_{\mu}{}^{\alpha} \, W_{\alpha\nu} + \frac{1}{4} \, W_{\alpha\beta} \, W^{\alpha\beta} \, g_{\mu\nu}
\end{equation*}
is the energy-momentum tensor of the vector field, $T_{\mu\nu}^{m}$
 is the energy-momentum tensor of matter, 
 \begin{equation}
 \lambda g_{\mu\nu} = \kappa\langle T_{\mu\nu}^{m} \rangle_{vac} = \kappa\rho_{vac}g_{\mu\nu},
 \end{equation}
 and
  \begin{widetext}
\begin{equation}
\tau_{\mu\nu}= V_{, \mu \, ;\nu} - \Box V(\Phi)  g_{\mu\nu} + \frac{1}{2} H(\Phi) \Phi_{\mu} \Phi_{\nu} - \frac{1}{4} H(\Phi) \Phi_{\alpha}\Phi^{\alpha} g_{\mu\nu} + \Omega(\Phi) W_{\mu}  W_{\nu}.
\end{equation}
 \end{widetext}

%\vspace*{0.4cm}

\begin{lemma}

Let us consider a vector field with dynamics defined by the Lagrangian (\ref{1}) and such that the functions $ V $ and $ H$ are given by
\begin{subequations}
\begin{equation}
V(\Phi) = \beta\Phi,  %\,c^2
\end{equation}
%$$ H(\Phi) = - \, \frac{3\, \beta^2 \, \kappa\, c^4}{(1 + \beta\,c^2 \, \kappa \,\Phi)} $$ 
\begin{equation}
H(\Phi) = - \frac{3 \beta^2 \kappa}{(1 + \beta \kappa \Phi)},
\end{equation}
\end{subequations}
where $ \beta$ are dimensionless constants, one for each vector field.
Then, the dynamical equation (\ref{13625}) takes the form
  \begin{equation}
  W^{\alpha\nu}{}_{; \nu} + 2 \beta (\kappa T_{m} + 2 \lambda) W^{\alpha} = 0.
  \label{13621}
  \end{equation}
 where $T_{m} = T_{\mu}^{m\mu}$ is the trace of the energy-momentum tensor of matter.
 
 \end{lemma}
 
We can now state the following theorem that establishes a gravitational mass generation mechanism for vector fields:

\begin{theorem} 
 
  When $ T_{m}$  vanishes and $\lambda \neq 0$,  the vector field acquires a mass term, one for each vector field according to the value of the specific non-minimal curvature-matter coupling constant $ \beta$, and the dynamical equation for the field $ W^{\mu\nu}$ becomes
\begin{equation}
W^{\alpha\nu}{}_{; \nu}  + 4 \beta \lambda W^{\alpha} = 0.
 \label{13622}
\end{equation} 

 \end{theorem}

\begin{corollary} 
 
 In the case where the matter energy-momentum tensor is traceless and $ \lambda$ vanishes, the field $ W^{\mu\nu}$ satisfies the dynamical equation
\begin{equation}
W^{\alpha\nu}{}_{; \nu} = 0. \label{13619}
\end{equation}
 
 \end{corollary}
 
 Note that, although equation (\ref{13619}) has the same form as ordinary Maxwell's equation in curved space-time, the metric field contained in the covariant derivative $\nabla_{\mu}$ is the solution of the modified field equation (\ref{13618}), which is not gauge-invariant.

We summarise the main features of the gravitational mass generation process for vector bosons:

\begin{itemize}
\item{The free $ W^{\mu}$ field obeys Maxwell\rq s dynamics defined by the Lagrangian $ L = - \, 1/4 \, W_{\mu\nu} W^{\mu\nu} $;}
\item{ There is a non-minimal coupling of the field with gravity controlled by the scalar $ \Phi$ coupled to the gravitational metric;}
\item{The state of matter is represented by a traceless energy-momentum tensor and the constant $\lambda$;}
\item{As a consequence of this coupling, the vector field acquires a mass;}
\item{The value of the mass does not depend on Newton's gravitational constant.}
\end{itemize}

The existence of massless vector fields in Nature is understood in the present framework as meaning that only the photon and the gluons couple minimally to gravity. The dynamics of all other vector fields is defined by the Lagrangian (\ref{1}).

The vacuum energy density plays a key role in the present mechanism and could, in principle, be used to constrain the values of the couplings $\beta$. Unfortunately, there is no reliable estimate of the vacuum energy contribution from the SM fields, which depends on the cut-off scale one adopts and ignores the influence of gravity.
For a cut-off of the order of the Planck mass, a standard calculation yields the value $\rho_{vac} \sim 10^{71}\, \mbox{GeV}^4$, which implies $\lambda \sim 10^{32}\, \mbox{GeV}^2$. In this case, taking $m_W \approx 80$ GeV as the mass term of the Proca equation (\ref{13622}), we obtain
\[
\beta_{|\mbox{\footnotesize{Planck scale}}} = \frac{m_W^2}{4\lambda} \sim 10^{-30}.
\]
 On the other hand, at the SM scale the vacuum energy density is of the order $\rho_{vac} \sim 10^{8}\, \mbox{GeV}^4$ \cite{MARTIN_2012, SOLA_2013}, which on its turn implies $\lambda \sim 10^{-30}\, \mbox{GeV}^2$. In this case, for the same mass of the $W$ boson, we obtain $\beta_{|\mbox{\footnotesize{SM scale}}} \sim 10^{33}$.
 Finally, for the observational value of $\rho_{vac} \sim 10^{-47}\, \mbox{GeV}^4$ \cite{PLANCK_2020}, we get $\beta_{|\mbox{\footnotesize{observ.}}} \sim 10^{88}$.

\section{A redundant Lagrangian for Dirac\rq s dynamics}

 As in the previous case, we may consider a fermion field $ \Psi$ coupled to gravity with dynamics defined by the Lagrangian
%  \begin{widetext}
 \begin{equation}
 L = L_{D} + \frac{1}{\kappa} R + \frac{1}{\kappa}S(N) R + \frac{1}{2\kappa} \, B(N) N^{\mu} N_{\mu} + L_{m} 
 \label{2}
 \end{equation}
 %\end{widetext}
where  $ L_{D}$ is Dirac\rq s Lagrangian, $ N$  is given by  $ N = \bar{\Psi} \Psi$, and $ L_{m}$ is the matter Lagrangian. 
Variation of $ g_{\mu\nu}$ yields
% \begin{widetext}
\begin{equation}
\frac{1}{\kappa} \left(1 + S\right)  \left( R_{\mu\nu} - \frac{1}{2} R g_{\mu\nu} \right) = - T_{\mu\nu}^{D} - \tau_{\mu\nu} - T_{\mu\nu}^{m} 
\label{23618}
\end{equation}
% \end{widetext}
where $T_{\mu\nu}^{D}$ is the energy-momentum tensor of the fermion field
$$ 
T_{\mu\nu}^{D} = \frac{1}{4} \left( \bar{\Psi} \gamma_{\mu} \nabla_{\nu} \Psi - \nabla_{\mu} \bar{\Psi} \gamma_{\nu} \Psi \right) + \mbox{symm}(\mu \leftrightarrow\nu),$$
 $T_{\mu\nu}^{m}$ is the energy-momentum tensor of matter, and
  %\begin{widetext}
  \begin{equation}
   \kappa\tau_{\mu\nu}= S_{, \mu \, ;\nu} - \Box S g_{\mu\nu} +  B N_{\mu} N_{\nu} - \frac{1}{2} B  N_{\alpha} N^{\alpha} g_{\mu\nu}. 
  \end{equation}
 %\end{widetext}
Using the trace of equation (\ref{23618}), the dynamical equation for $ \Psi$ takes the form\begin{equation}
i \gamma^{\mu} \nabla_{\mu} \Psi +  \Theta \Psi = 0,
\label{23625}
\end{equation}
where
 $ \Theta = R S' -  B' N_{\alpha} N^{\alpha} - B \Box N $, and a prime means derivative with respect to $ N.$  
 
 Following the same steps as in the previous section, we can now state the following theorem that establishes a mass generation mechanism for a spinor field:

%\vspace*{0.4cm}

\begin{theorem}

 Let us consider a spinor field controlled by the Lagrangian (\ref{2}) and such that the functions $ S $ and $ B $ are given by 
\begin{subequations}
\begin{equation}
S(N) = a + b N,
\end{equation}
\begin{equation}
B(N) = \frac{3 b^{2}}{(1 - a - b N)},
\end{equation}
\end{subequations} 
 where $ a $ and $ b $ are constants. In this case, the dynamical equation for $ \Psi$ becomes
  \begin{equation}
  i \gamma^{\mu} \nabla_{\mu} \Psi  -  m %c \, 
  \Psi = 0,
  \label{23621}
  \end{equation}
where the mass depends only on constants $a, b$ and the trace of $T_{\mu\nu}^m$:
%$$ m = \frac{b}{(1 - a)c^2} \, T_m .$$
\begin{equation}
m = \frac{b}{(1 - a)} T_m .
\end{equation}

\end{theorem}

\begin{corollary} 
 
 In the case the matter energy-momentum tensor is traceless, the equations (\ref{23625}) and (\ref{23618}) imply that the field $ \Psi$ satisfies Dirac's dynamics
\begin{equation}
 i \gamma^{\mu} \nabla_{\mu} \Psi = 0.
\end{equation}

\end{corollary}

Let us summarize the main features of the gravitational mass generation mechanism for fermions:
\begin{itemize}
\item{The free $ \Psi$ field obeys Dirac\rq s dynamics driven by the Lagrangian $ L_{D} $;}
\item{ There is a non-minimal coupling of the field with gravity controlled by the scalar $ N = \bar{\Psi} \Psi $ coupled to the gravitational metric;}
\item{The state of matter is represented by an energy-momentum tensor with constant trace;}
\item{As a consequence of this coupling the spinor field acquires a mass;}
\item{The value of the mass does not depend on Newton's gravitational constant.}
\end{itemize}

\section{Final Remarks}

Let us point out that in the BEH mechanism the mass appears as a consequence of the vacuum energy of the Higgs field, represented by an energy-momentum tensor with the form $ T_{\mu\nu} = U_0 \, g_{\mu\nu} $, where the constant $ U_0$ is given by the extremum of a potential $ U.$ This implies that, at the classical level, the BEH mechanism can be considered a particular case of the present mechanism. We also note that although in the BEH mechanism the Higgs field must be in direct interaction with the vector field, this is not necessary for the present mechanism, as the mass appears through the mediation of the gravitational field. 
In other words, gravity is the catalyst of the process of mass generation.

It should be noted that non-minimal curvature-matter couplings lead to violations of the Strong Equivalence Principle (SEP) that could, in principle, be detected in strong field regimes. Although the SEP is strongly constrained at the solar system scale \cite{WILL_2018}, constraints at larger scales are, however, poorly studied \cite{BONVIN_2018}.

 \subsection*{Acknowledgements}
We would like to acknowledge the financial support from Brazilian agencies Faperj and CNPq.

\end{document}